\documentclass[amssymb,twocolumn,10pt,prl]{revtex4}
\usepackage{dcolumn}
\usepackage{bm}
\usepackage{amsmath}

\begin{document} 
\noindent
{\large\bf Comment on ``Experimental Realization of a Three-Qubit Entangled \textit{W} state"} \\

Recently Eibl \textit{et al.} \cite{EKBKW} reported the experimental observation of the three-photon polarization-entangled \textit{W\/} state \cite{DVC} using spontaneous parametric down-conversion. Conditioned on the detection of one photon in each of the four modes \textit{a}, \textit{b}, \textit{c}, and \textit{t} (see Fig.~1 of \cite{EKBKW}), the observed three-photon \textit{W\/} state in modes \textit{a}, \textit{b}, and \textit{c}, is
\begin{equation}
|W\rangle = \frac{1}{\sqrt{3}}(|HHV\rangle_{abc} + |HVH\rangle_{abc} + |VHH\rangle_{abc}) .
\end{equation}
It is well known that the pure state (1) represents one of the two classes of irreducible tripartite entanglement \cite{DVC}. In order to experimentally characterize the three-particle entanglement of the observed state the authors of \cite{EKBKW} analyzed the correlations between the measurement results in the three modes $j=a,b,c$, of the linear polarization operators $\hat{\sigma}_j =|\phi^{+}_{j}\rangle\langle \phi^{+}_{j}|-|\phi^{-}_{j}\rangle\langle \phi^{-}_{j}|$, where $|\phi^{\pm}_{j}\rangle = (1/\sqrt{2})(|R\rangle\pm e^{i\phi_j}|L\rangle)$. Specifically, they tested the Bell-type Mermin inequality \cite{Mermin90}, $S_M = E(\phi_a,\phi_b,\phi^{\prime}_{c})+E(\phi_a,\phi^{\prime}_{b},\phi_c)+E(\phi^{\prime}_{a},\phi_b,\phi_c)-E(\phi^{\prime}_{a},\phi^{\prime}_{b},\phi^{\prime}_{c}) \leq 2$, for the analyzer settings $\phi_j = \pi/2$ and $\phi^{\prime}_{j}=0$. The aim of this Comment is to show that, actually, the particular measurements involved in the experiment testing the Bell inequality $S_M \leq 2$ cannot be used to prove the existence of genuinely quantal tripartite correlations in the observed state, and therefore this state cannot conclusively be identified (within the experimental errors) with the \textit{W\/} state.

In the first place it should be noted that Mermin's inequality does not discriminate the purely quantum-mechanical correlation due to three-particle entanglement from the hybrid local-nonlocal hidden variables models \cite{Svetlichny,MPR,CGPRS} that assume nonlocal correlation between two of the three particles, but only \textit{local\/} correlation between these two particles and the third one. Indeed, even a maximal violation of Mermin's inequality, $S_M = 4$, can be easily reproduced by such a model \cite{CGPRS,Cereceda}. In order to experimentally rule out the hypothesis of partial nonlocality and to detect full three-particle entanglement, Svetlichny inequality (SI) \cite{Svetlichny}, $S_V \leq 4$, has to be invoked, where
\begin{align}
S_V = & \,\, E(\phi_a,\phi_b,\phi_c)+E(\phi_a,\phi_b,\phi^{\prime}_{c})
+E(\phi_a,\phi^{\prime}_{b},\phi_c)  \nonumber  \\
& +E(\phi^{\prime}_{a},\phi_b,\phi_c) - E(\phi_a,\phi^{\prime}_{b},\phi^{\prime}_{c})
- E(\phi^{\prime}_{a},\phi_b,\phi^{\prime}_{c})  \nonumber   \\
& - E(\phi^{\prime}_{a},\phi^{\prime}_{b},\phi_c) - E(\phi^{\prime}_{a},\phi^{\prime}_{b},\phi^{\prime}_{c}).
\end{align}
In fact, if the measured value of $S_V$ for the observed state turns out to be greater than 4, then the three-particle entanglement feature and the three-particle nonlocality (or nonseparability) of the given quantum state would have been unambiguously demonstrated. However, the measurements carried out in \cite{EKBKW} do not allow SI to be violated, nor even in the case of ideal experimental conditions. Indeed, for the analyzer settings $\phi_j = \pi/2$ and $\phi^{\prime}_{j}=0$, the quantum prediction of $S_V$ for the (ideal) pure state (1) is equal to $S^{\text{QM}}_{V}= 3$. Hence, as $S^{\text{QM}}_{V} < 4$, it is concluded that the set of correlation measurements performed in \cite{EKBKW} cannot be used for the verification of the existence of both tripartite entanglement and full nonseparability of the observed state, and therefore experimental observation of the three-photon polarization-entangled \textit{W\/} state cannot still be conclusively established. We note that a similar situation to that described in this Comment already occurred in some of the first experiments designed to produce the maximally entangled three-particle GHZ state \cite{MPR,SU02}.

Finally, let us mention that the maximum value of $S_V$ predicted by quantum mechanics for the \textit{W\/} state is $S^{\text{max}}_{V}(W)= 4.354$ \cite{Cereceda}, which is realized, for instance, by choosing $\phi_j = 35.264^{\circ}$ and $\phi^{\prime}_{j}=144.736^{\circ}$. Therefore, in order to optimally confirm that the state produced indeed exhibits the required features of full entanglement and full nonseparability, the correlation experiment in \cite{EKBKW} ought to test Svetlichny inequality, with the analyzer settings adjusted at $\phi_j \simeq 35.26^{\circ}$ and $\phi^{\prime}_{j} \simeq 144.74^{\circ}$, instead of $\phi_j = \pi/2$ and $\phi^{\prime}_{j}=0$.

\vspace{.3cm}

\noindent Jos\'{e} L. Cereceda (jl.cereceda@telefonica.net) \\
C/Alto del Le\'{o}n 8, 4A, 28038 Madrid, Spain   \\

\end{document}